\documentclass[a4paper,pdftex,reqno,11pt]{amsart}
\normalfont\sffamily
\allowdisplaybreaks[1]

\usepackage{geometry}
\usepackage{amsfonts}
\usepackage{graphicx}
\usepackage{enumerate}
\usepackage{courier}
\usepackage{times}
\usepackage{amsmath,amssymb}
\usepackage{geometry}
\usepackage{hyperref}

\theoremstyle{plain}
\newtheorem{Theorem}{Theorem}[section]

\theoremstyle{definition}

\theoremstyle{remark}

\begin{document}


\title[Which criteria qualify power indices for applications?]{Which criteria qualify power indices for applications?\\ A comment to {\lq\lq}The story of the poor Public Good index{\rq\rq}}

\author{Sascha Kurz}
\address{Sascha Kurz, University of Bayreuth, 95440 Bayreuth, Germany}
\email{sascha.kurz@uni-bayreuth.de}

\date{}

\abstract{We discuss possible criteria that may qualify or disqualify power indices for applications. Instead of providing final answers we merely ask 
questions that are relevant from our point of view and summarize some material from the literature.\\[2mm]
}}

\maketitle

\section{Introduction}
\label{sec_intro}
Consider a committee that draws decisions via weighted voting. Given a heterogeneous committee, i.e., not all members have the same weight, 
the question for the relative distribution of power or influence within the committee naturally arises. The relative weight distribution 
comes first to mind. However, it is a very poor proxy in at least some situations. Consider e.g.\ a committee with three members drawing 
decisions by simple majority, where the weight distribution is given by $(47~\%,36~\%,17~\%)$. The first two committee members seem to have 
a larger influence on the committee's decision than the third member. However, any two (or three) members can jointly implement a proposal. 
Thus, all three committee members have arguably the very same influence on the group decision. The underlying mathematical reason for 
the poor approximation of power or influence by weights lies in the variance of the possible weight vectors. I.e., our example might also be 
captured by weight vectors like  e.g.\ $(2~\%,49~\%,49~\%)$, $(30~\%,40~\%,30~\%)$, or $(40~\%,15~\%,45~\%)$. Indeed the set of weights 
representing such a weighted game is a full-dimensional polytope, see e.g.\ \cite{kaniovski2018representation}.  

So, there is a need for measurements of power or influence that do not depend on a given weighted representation but on the underlying \textit{game}. 
(We will be more precise in the next section.) To this end so-called \emph{power indices} were invented, see e.g.\ \cite{riker1986first} for a discussion about 
the introduction of the first power index. We remark that there is some discussion about what \emph{power} precisely means, see e.g.\ \cite{riker1964some} and 
the citations thereof. However, we will not go into that here. The issue at stake is that several variants of power indices were proposed so far and still are, 
see e.g.~\cite{casajus2018coleman},  
so that one needs criteria to decide which power index to use. The discussion is quite old, see e.g.\ \cite{holler1983power} and the references therein. The 
current occasion is a comment of Manfred Holler \cite{hollerstory}, on a conjecture of Reinhard Selten, arguing that in some cases the choice or the exclusion of a certain power index is related to 
the fame and influence of its proposer. Of course I agree that this should not be a relevant criterion and should be prevented in science. Here I would like to 
comment on the Public Good index but mainly I would like to use the possibility to stimulate a broader discussion on reasonable criteria that qualify a power index 
for a given application. As this paper is also a comment and not a pure research paper, it contains many personal opinions and the selection of topics and 
literature is rather biased.

The remaining part of this comment is structured as follows. The necessary preliminaries on simple and weighted games as well as on power indices 
are given in Section~\ref{sec_preliminaries}. The main discussion on selection criteria for power indices is located in Section~\ref{sec_selection}. 
The problem of whether we are still missing some important indices is addressed in Section~\ref{sec_missed}. We draw a conclusion in Section~\ref{sec_conclusion}.

\section{Preliminaries}
\label{sec_preliminaries}
For a positive integer $n$ let $N=\{1,\dots,n\}$ be the set of players.  A \emph{simple game} is a mapping $v\colon 2^N\to\{0,1\}$ from  the  subsets  of
$N$ to  binary  outcomes  satisfying $v(\emptyset)=0$, $v(N)=1$, and $v(S)\le v(T)$ for all $\emptyset \subseteq S\subseteq T\subseteq N$. Here, a subset
$S\subseteq N$ is considered as the set of players that are in favor of a proposal, i.e., which vote {\lq\lq}yes{\rq\rq}. If $v(S) = 1$ we call coalition
$S$ \emph{winning} and \emph{losing} otherwise. Given a simple game, by $\mathcal{W}$ we denote the set of winning and by $\mathcal{L}$ the set of losing 
coalitions. If a coalition $S$ is winning but each proper subset is losing, then we call $S$ \emph{minimal winning}.  Similarly, if $S$ is losing but each proper 
superset of $S$ is winning, then we call $S$ \emph{maximal losing}. If $v(S)=v(S\cup\{i\})$ for all $S\subseteq N$, then we call player $i$ a \emph{null player}. 
Two players $i$ and $j$ are \emph{equivalent} if $v(S\cup\{i\})=v(S\cup\{j\})$ for all $S\subseteq N\backslash\{i,j\}$. A simple  game $v$ is  called
\emph{weighted} if  there  exist  \emph{weights} $w\in\mathbb{R}_{\ge 0}^n$ and a \emph{quota} $q\in\mathbb{R}_{>0}$ such that $v(S) = 1$ if and only if
$w(S):=\sum_{i\in S} w_i\ge q$. As notation we use $v=[q;w_1,\dots,w_n]$ for a weighted game. An example of a non-weighted simple game $v$ is given by a 
family consisting of the parents $\{F,M\}$ and two kids $\{A,B\}$ jointly 
deciding on the next weekend trip. Assume that a proposal is accepted if at least one member of every group, i.e., parents and children, agrees. Since 
$\{F,M\}$ and $\{A,B\}$ are two losing coalitions but $\{F,A\}$ and $\{M,B\}$ are two winning coalitions consisting of the same set of players, the underlying 
simple game is not weighted. In our example players $F$ and $M$ as well as players $A$ and $B$ are equivalent. The minimal winning coalitions are given by 
$\{F,A\}$, $\{F,B\}$, $\{M,A\}$, and $\{M,B\}$. The \emph{Public Good index} $\operatorname{PGI}(v)_i$ of a player $i\in N$ is given by the number of minimal 
winning coalitions that contain player~$i$ divided by the total number of players in all minimal winning coalitions. I.e., in our example the Public Good index 
assigns equal influence to all four players.

The Public Good index is just one example of a \emph{power index}. Here, we understand by a power index a parametric family of mappings from the set of simple games 
on $n$ players\footnote{It is of course also possible to define a power index just on special subclasses of simple games like e.g.\ weighted games. Examples 
for the latter are given by the power indices based on weighted representations from \cite{kaniovski2015average}.} into $\mathbb{R}^n$, i.e., to each player of a 
simple game we assign a real number. By that definition we allow a large variety of possible power measures but also exclude constructions that depend on the 
representation of a (weighted) simple game like e.g.\ the Colomer index \cite{colomer1995paradox}. Instead of restricting the class of power indices, we introduce 
properties of power indices and argue that it is quite natural to expect some of these properties to be satisfied for a \textit{reasonable} power index. We call 
a power index $g$ \emph{positive} if $g_i(v)\ge 0$ and $g(v)\neq 0$ for every simple game $v$ and all players $i$ in $v$. If $\sum_{i\in N} g_i(v)=1$ for every 
simple game $v$, then $g$ is called \emph{efficient}. In our context efficiency is a relevant property if we want to measure the \emph{relative} distribution 
of power or influence. However, given any positive power index $\tilde{g}$ we can construct an efficient version via normalization, i.e., $g_i(v):=\tilde{g}_i(v)/\sum_{j\in N} \tilde{g}_j(v)$, 
see e.g.~\cite[Lemma 3.4]{kurz2016inverse}. We say that a power index $g$ has the \emph{null player property} if we have $g_i(v)=0$ for every null player $i$ 
in an arbitrary simple game $v$. For any bijection $\tau\colon N\to N$ we write $\tau v$ for the simple game given by $\tau v(S)=v(\tau(S))$ for any coalition 
$S\subseteq N$. With this, we call a power index $g$ \emph{symmetric} if we have $g_{\tau(i)}(\tau v)=g_i(v)$ for every simple game $v$, every player $i\in N$, and every 
permutation $\tau$ of $N$. If we always have $g_i([q;w_1,\dots,w_n])\ge g_j([q;w_1,\dots,w_n])$ for all indices $i,j$ with $w_i\ge w_j$, we speak of \emph{local 
monotonicity}. In order to define local monotonicity for non-weighted games, we need a bit more notation. Let $v$ be an arbitrary simple game. If $v(S\cup\{i\})\ge 
v(S\cup\{j\})$ for all $S\subseteq N\backslash\{i,j\}$, we write $i\succeq j$. With this, we call a power index $g$ locally monotonic for $v$ is $g_i(v)\ge g_j(v)$ 
whenever $i\succeq j$. If the relation $\succeq$ is a complete, i.e., if we always have $i\succeq j$ or $j\succeq i$, we call $v$ \emph{complete simple game}. With this, 
a power index $g$ is locally monotonic if it is locally monotonic for all complete simple games.\footnote{Of course this definition may be restricted to the class of 
weighted games.}

A \emph{TU game} is a mapping $v\colon 2^N\to\mathbb{R}$, i.e., each coalition is associated with a real number. Several power indices have a generalization to TU games. In that 
case one speaks of a \emph{value}. We remark that e.g.\ the Shapley value for TU games was invented first and later specialized to the Shapley-Shubik index for simple 
games.      

\section{Selection criteria for power indices}
\label{sec_selection}

In this section we would like to collect some criteria that may guide a user to select an appropriate power index for a given application. 

\subsection{The axiomatic approach}
\label{subsec_axiomatic}
The basic idea of the axiomatic approach is very appealing. In principle one looks for simple properties of a power index, called axioms. For a given practical 
problem then one can check, which axioms have to be satisfied. Since some of those axioms are violated by some power indices such an analysis can help to limit 
the set of \textit{reasonable} power indices for the given application. This kind of reasoning was e.g.\ be applied in \cite{napel2017responsibility} for the 
selection of value, for the allocation of cartel damages. 
That approach calls for a large list of easy to evaluate and well justified axioms. Of course it is helpful if the axioms are disjunctive enough so that one finally 
ends up with a rather small set of possibilities. Indeed, the ultimate goal is an axiomatization of a given power index, i.e., a set of inclusion-minimal axioms that uniquely 
characterizes a power index in the class of all theoretically possible power indices. Of course it is good to have a unique answer, but on the other hand this request 
ends up in a search for axioms that exactly do this job and hopefully can be motivated in a more or less reasonable way. The large literature on axiomatizations 
teaches us that it seems to be not that hard to come up with some axiomatization for a large variety of power indices. This is just a personal opinion, but in 
most axiomatizations in the literature there is one axiom that is not as convincing as the others and looks somehow \textit{constructed} for the purpose of 
getting an axiomatization. (Admittedly, there are exceptions like e.g.\ Youngs axiomatization of the Shapley value \cite{young1985monotonic}.) Conceptionally, starting 
from a given power index and searching for reasonable axioms seems not to be the right direction from our point of view. Moreover, I would say that the (temporary) 
absence of an axiomatization should not be a disqualifier for a reasonable power index. 

What about the other direction? Which axioms may be seen as rather essential for nearly every application? My personal list consists of positivity, symmetry, efficiency, 
and the null player property. Being a mathematician, I of course like linearity, i.e., $g(\alpha u+\beta v)=\alpha g(u)+\beta g(v)$, or the reformulation and restriction 
to the class of simple games. However, I am not convinced that linearity is inherent in every application of power indices. As pointed out in \cite{hollerstory}, the 
violation of local monotonicity was e.g.\ used in \cite{felsenthal1995postulates} to disqualify the Public Good index as a reasonable power index. As well as Holler, 
we do not agree on that point. My own answer on that issue is a bit indirect and goes as follows. If we say that symmetry, efficiency, the null player property and 
local monotonicity are essential for a reasonable power index, then have a look at the classes of power indices based on the representation polytope of weighted games 
in \cite{kaniovski2015average,kaniovski2018representation}. Either these indices are also reasonable power indices or we are missing further \textit{essential} axioms. 
Moreover, as a mathematician, I see little reason why the requested local monotonicity should be restricted to simple games. We can make the same kind of comparisons 
for complete simple games and, partially, also for simple games. Is it also natural to require local monotonicity it that stronger sense? Anyway, the violation of 
local monotonicity is just one example in a long list of possible voting paradoxes, see e.g.\ \cite{nurmi1999voting}.

\subsection{Qualitative properties}
\label{subsec_qualitative}

As mentioned, the possibility to verify axioms in a practical application is very appealing. If one weakens the requirements of a formal axiom one might look 
for some qualitative behavior. If predictions are made within a model based on a power index, different power indices may be evaluated using econometric analysis. 
As a continuation of \cite{kauppi2004determines} such a comparison was performed e.g.\ in \cite{zaporozhets2016key}. Apart from that, the idea of qualitative properties 
is rather vague, but I want to give two examples. Consider a game with many players like a publicly traded stock company. What can we say about the power distribution 
of the major stock holders? Let us look at the extreme case of one stock holder owning a relatively large share. Mathematical analysis of the possible power of the 
largest player in a weighted game gives a rather clear picture.
\begin{Theorem}(\cite[Theorem 8]{kurz2018power}) 
Let $g$ be the Nucleolus, the Public Good, the Deegan-Packel, the minimum sum representation, the average weight, or the average representation index, then for
each $\varepsilon>0$ there exists an integer $N(\varepsilon)$ such that either $g_i(v)=1$ or $g_i(v)\le\tfrac{1}{2}+\varepsilon$ for each simple game $v$ on  
$n\ge N(\varepsilon)$ players and an arbitrary player $1\le i\le n$.
\end{Theorem}
In contrast to that, the Shapley-Shubik, the Penrose-Banzhaf, and the Johnston index permit the existence of simple games such that $g_i(v)$ is arbitrarily close to any 
number in $\left[\tfrac{1}{2},1\right]$. 

In a practical application, do we have this latter {\lq\lq}continuum{\rq\rq} of possible power values for the largest player or is there a clear cut (between $1$ and 
$\tfrac{1}{2}$)? This is certainly a different qualitative behavior, that may be observed in practice?

A second example was initiated in \cite{alon2010inverse}. Suppose we have we have some power index $g$ and some weighted game $v$ with player set $N$. If there exists 
a subset $I\subseteq N$ of the players such that $\sum_{j\in N\backslash I} g_j(v)\le\varepsilon$ for some small $\varepsilon\in\mathbb{R}_{>0}$, does there always exist 
a simple game $\tilde{v}$ with player set $I$ such that
$$
  \Vert g(v)-g(\tilde{v})\Vert_1= \sum_{i\in I} \left| g_i(v)-g_i(\tilde{v})\right|+\sum_{i\in N\backslash I} \left| g_i(v)\right|\quad\le\quad f(|I|)\cdot\varepsilon,
$$
where $f$ is a general function just depending on the cardinality of $I$? As shown in \cite{alon2010inverse}, such a result exists for the Penrose-Banzhaf index. In words 
we may reformulate the technical statement as follows. If for a simple game on $n$ players the power is concentrated on just $k<n$ players, then there exists a $k$-player 
simple game whose power distribution does not deviate too much from the one of the original simple game. In \cite{kurz2016inverse} it was e.g.\ shown that also the Public 
Good index allows such an Alon-Edelman type result while e.g.\ the Johnston index does not. Again, we have some kind of qualitative behavior that is different for different 
power indices. Admittedly, the second example is significantly harder to grasp than the first one. 

Further such examples might be rather useful. In \cite{dubey1979mathematical} a lot of mathematical properties of the Penrose-Banzhaf index are mentioned. It might be 
worthwhile to compare them to similar results for other power indices. Starting with a purely mathematical motivation, one then has to check which of the phenomena 
can be verified in practical applications.

\subsection{Beauty and generalizability}
\label{subsec_beauty}

It is interesting to know if for a given power index a generalization to a value, i.e., an extension to the class of TU games, exists. This is e.g.\ the case for the 
Shapley-Shubik and the Penrose-Banzhaf index, as well as the Nucleolus, but not for the Public Good index. The first two mentioned power indices also allow a 
generalization to non-binary decisions -- either discrete or non-discrete, see e.g.\ \cite{kurz2014measuring,kurz2018importance}. Continuous or interval decisions 
are not completely uncommon. In the context of a {\lq\lq}fair{\rq\rq} allocation of voting weights for differently sized groups this model has been studied, e.g., 
in \cite{kurz2017democratic}. The existence of a generalization of a concept to a large class of structures may be seen as an indication that the concept is somehow 
meaningful. Unfortunately, this does not answer the question for \textit{what} purpose it is meaningful.  

If we consider weighted games with many players where just a few have a non-vanishing weight, then some power indices allow so-called \emph{limit results}. I.e., if 
the number of minor players increases but their total weight sum is fixed and the relative weights go to zero, then there is a stable power distribution of the major 
players. As shown in \cite[Section 10]{dubey1979mathematical} such kind of a limit result exists for the Penrose-Banzhaf index only if further assumptions on the weight 
ratios of the minor players are made. For the Shapley-Shubik index such an assumption is not necessary, which is the more {\lq\lq}beautiful{\rq\rq} result. For the 
nucleolus there exists a very simple and explicit upper bound on the deviation between power and weights for finite weighted games, see \cite[Lemma 1]{kurz2014nucleolus}. 
Compared to the known situation for the Shapley-Shubik index, see \cite{neyman1982renewal}, we would call this more {\lq\lq}beautiful{\rq\rq}. However, this depends on 
our current state of knowledge and the argument of beautiful theories has failed badly in e.g.\ physics \cite{brumfiel2011beautiful}.

\subsection{Undesirable criteria}
\label{subsec_undesirable}
We completely agree with \cite{hollerstory} that the fame and influence of the proposer of a power index should not have a significant impact on its usage. However, 
we have nothing to say whether that happened to the Public Good index. We go a step further. When selecting the {\lq\lq}best{\rq\rq} index, we should also not be 
limited to the already published power indices. In principle it may be the case that we have missed the best power index for some applications so far. We will 
try to address this issue in Section~\ref{sec_missed}. 

As argued in \cite[Section 4]{kurz2015mostly} the availability of easy-to-use software packages to compute power indices can limit their application, e.g., for 
the nucleolus. While this is understandable from a practical point of view, this should not be a desireable selection criterion for the choice of an appropriate 
power index.

\section{Have we missed an index?}
\label{sec_missed}
It seems natural that the selection of the best alternative may be limited to the set of the known alternatives. But this is not the case. New alternatives may be 
invented or even determined using optimization techniques. In this section we would like to address the question whether we have missed an index and also give a further 
justification for the Public Good index.

\subsection{Systematization of power indices}
\label{subsec_systematization}
Since a large variety of power indices was introduced in the literature so far, it makes sense to try to catalogue them in a systematic way. Here we will stick 
to the systematization from \cite{kurz2016inverse}, where all the subsequently occurring power indices are defined and cited. One idea is to count coalitions of 
a certain type $\mathcal{T}$, i.e., 
$$
  S^{\mathcal{T}}_i(v)=\#\left\{U\in\mathcal{T} \,:\, \{i\}\subseteq U\subseteq N \right\}, 
$$
c.f.~\cite[Subsection 3.5]{kurz2016inverse}. Afterwards we can normalize via $S^{\mathcal{T}}_i(v)/\sum_{j\in N} S^{\mathcal{T}}_j(v)$ to an efficient version, so that scaling factors are irrelevant. So, 
ignoring scaling factors, the types of winning, swing or critical, and minimal winning coalitions lead to the \emph{K\"onig-Br\"auninger}\footnote{The K\"onig-Br\"auninger index is 
proportional to the Public Help and the Chow parameter index and also known under the name inclusiveness and Zipke index.}, the Penrose-Banzhaf\footnote{For the Penrose-Banzhaf 
index we have the technical detail that a critical coalition is only counted for those contained players that are critical, i.e., $v(U)-v(U\backslash\{i\})=1$.}, and the Public 
Good index, respectively. For complete simple games the tightening of a minimal winning coalition is called shift-minimal winning coalition and the corresponding power index based 
on $S^{\mathcal{T}}$ is called \emph{Shift index}. Are there other types of coalitions that have a relation to power? Removing the Public Good index from the above list, c.f.\ 
\cite{nurmi1997power}, would certainly leave some gap.

In the above construction a special coalition $U\in\mathcal{T}$ may be counted for several players. Another variant is to distribute the contribution of a coalition 
equally among the involved players. This concept is called \emph{equal division} transform in \cite[Definition 3.22]{kurz2016inverse} and turns e.g.\ the Penrose-Banzhaf and the 
Public Good index into the \emph{Johnston} and the \emph{Deegan-Packel index}, respectively. For the Shift index the corresponding equal division variant is called 
\emph{Shift-Deegan-Packel index}. For the K\"onig-Br\"auninger index we are not aware of a published name of the corresponding equal division variant.     

The Shapley-Shubik index arises similarly as the Penrose-Banzhaf index from counting swing coalitions. Additionally, the contribution of every swing coalition is 
weighted by some factor solely depending on the cardinality of that coalition. Those weighted versions might of course also be studied for the other mentioned types of 
{\lq\lq}special{\rq\rq} coalitions.

\subsection{Design of new power indices}
\label{subsec_design}
In \cite{widgren2001probabilistic} Widgr\'en relates the normalized Penrose-Banzhaf index and the Public Good index via $\operatorname{PBI}_i=(1-\pi)\operatorname{PGI}_i+
\pi\varepsilon_i$, see also \cite{hollerstory} for details and interpretation. Since the Penrose-Banzhaf index is locally monotonic, this relation can point to elements 
which are responsible for the PGI's violation of local monotonicity. This fact has motivated the research in \cite{freixas2016cost}, where the authors, e.g., considered 
convex combinations of the Penrose-Banzhaf and the Public Good index and asked for the minimum possible weight of the Penrose-Banzhaf index that yields a power index 
that is locally monotonic. From a general point of view, this is just an instance of an attempt to design a new power index with some desired properties. The restriction 
to convex combinations of established power indices reduces the complexity in the computational part and allows first analyses. However, the ultimate goal would 
be to design power indices in the entire class of all theoretically possible power indices just by prescribing desired properties. Currently, the list of 
broadly accepted properties is too short to end up with a manageable (parametric) class of power indices. In any case, integer linear programming techniques, as 
used in \cite{freixas2016cost}, seem to be very advantageous for the design aim.   

\section{Conclusion}
\label{sec_conclusion}
 We have argued that the Public Good index should not be excluded from the list of reasonable power indices for several reasons. We completely agree with \cite{hollerstory}  
 that the fame and influence of the proposer of a power index should not have a significant impact on its usage. We partially agree that the violation of local 
 monotonicity of the Public Good index is not a bug it's a feature, i.e., the $\operatorname{PGI}$ may detect special games. However, we have so far not seen any convincing 
 explanation of \textit{what} is detected. More analyses are needed in that direction. Moreover, the Public Good index naturally occurs in a systematic treatment 
 of known power indices. 
 
 Instead of providing answers, here we want to broaden the discussion on which criteria should qualify a reasonable power index. We have tried to collect some possible aspects 
 and would like to see more discussion on that question. Especially, we would highly welcome any suggestions for desireable properties of power indices that may further localize 
 the class of all {\lq\lq}reasonable{\rq\rq} power indices.


\end{document}